\documentclass[aps,prl,preprint,groupedaddress, citeautoscript]{revtex4-1}

\usepackage{soul}
\usepackage{color}
\setcitestyle{super}
\usepackage{graphicx}

\begin{document}
\title{\emph{In-situ} real-space imaging of crystal surface reconstruction dynamics via electron microscopy}


\author{Weizong Xu}
\author{Preston C. Bowes}
\author{Everett D. Grimley}
\author{Douglas L. Irving}
\author{James M. LeBeau}
\affiliation{Department of Materials Science and Engineering, North Carolina State University, Raleigh, NC 27695, USA}

\begin{abstract}

Crystal surfaces are sensitive to the surrounding environment, where atoms left with broken bonds reconstruct to minimize surface energy.
In many cases, the surface can exhibit chemical properties unique from the bulk.
These differences are important as they control reactions \cite{Surface_reaction} and mediate thin film growth \cite{Oxygen_conduct_science,2D_gas_Nature2011}.
This is particularly true for complex oxides where certain terminating crystal planes are polar and have a net dipole moment.
For polar terminations, reconstruction of atoms on the surface is the central mechanism to avoid the so called ``polar catastrophe'' \cite{Review_polar2000}.
This adds to the complexity of the reconstruction where charge polarization and stoichiometry govern the final surface in addition to standard thermodynamic parameters such as temperature and partial pressure.
Therefore, \emph{in-situ} atomic scale observations at the environmental conditions where these surfaces occur are essential to understand governing reconstruction mechanisms.
Here we present direct, \emph{in-situ} determination of polar SrTiO$_3$ (110) surfaces at temperatures up to 900 $^\circ$C using cross-sectional aberration corrected scanning transmission electron microscopy (STEM).
Under these conditions, we observe the coexistence of various surface structures that change as a function of temperature.
As the specimen temperature is lowered, the reconstructed surface evolves due to thermal mismatch with the substrate.
Periodic defects, similar to dislocations, are found in these surface structures and act to relieve stress due to mismatch.
Combining STEM observations and electron spectroscopy with density functional theory, we find a combination of lattice misfit and charge compensation for stabilization.
Beyond the characterization of these complex reconstructions, we have developed a general framework that opens a new pathway to simultaneously investigate the surface and near surface regions of single crystals as a function of environment.

\end{abstract}

\maketitle

Understanding the nature of SrTiO$_3$ surface reconstructions has many technologically important applications including catalysis \cite{SrTiO3_catalyst}, functional thin film growth \cite{Oxygen_conduct_science}, and its use in electronic devices \cite{2D_gas_Nature2011, Stemmer_Mobility}.
In many ways, SrTiO$_3$ serves as a model material to investigate single crystal complex oxide surface reconstructions.
A rich set of possible surface structures are observed with many factors driving the relative stability of different reconstructions.
These include Ti being able to change charge state, changes in the surface stoichiometry \cite{Wang_PRB_2011, Wang_PRB_2014},  as well as the influence of temperature and partial pressure of relevant species.
Many studies have focused on determination of the structure of the (100) surface. 
As a result, many models have been proposed for reconstructed non-polar (100) surfaces \cite{Castell_review, LD_Marks_nature_2002} while work is just emerging on the polar (110) habit \cite{Wang2016,Wang_PRL_2013,LD_Marks_nature_2010}.
The (110) surface can terminate with a plane of SrTiO$^{4+}$ or O$_2^{4-}$. 
If unmodified in either stoichiometry or oxidation state, both preserve conditions of a divergent electrostatic potential that would ultimately lead to the ``polar catastrophe'' \cite{Pojani1999179, Russell_PRB2008}.

Characterization of these surfaces at high temperatures remains challenging.
While scanning tunneling microscopy (STM) has been used to investigate kinetics of the surfaces at high temperatures the resolution quickly degrades \cite{PRL_STM_791K, HOT_STEM_2006}.
Beyond STM, electron microscopy, whether imaging or diffraction, has a rich history of aiding in the determination of crystal surfaces including nanoparticles and single crystals \cite{Ebeam_diffraction}.
Due to sample holder instability, however, these studies were conducted at room temperature where surface contamination becomes a challenge.

Recently, \emph{in-situ} electron microscopy has undergone a renaissance with the introduction of membrane-based heating devices that minimize thermal drift of the specimen \cite{Protochips_2009, Allard_2015}.
This breakthrough has enabled atomic scale investigation of high temperature thermodynamic and kinetic processes.
Thus, there is considerable potential for deciphering many of the complexities associated with surface reconstructions, such as the evolution of defects at elevated temperatures \cite{HOT_STEM_2006}.
\emph{In-situ} electron microscopy of single crystals, however, has been limited by preparation methods that leave contaminants on the sample surface, such as with focused ion beam \cite{FIB_Damage}.

In this Letter, we report SrTiO$_3$ (110) surface reconstructions determined directly  at temperatures up to 900$^\circ$C using cross-sectional aberration corrected scanning transmission electron microscopy (STEM).
Since it is critical that the surface be free of contamination to the greatest extent possible for the study of single crystal surfaces, we develop and apply a novel preparation approach.
Then a combination of atomic resolution STEM imaging and spectroscopy is used to measure the position of the atoms, including cations and oxygen at and just below the surface, as a function of temperature.
The surface and near surface composition and electronic structure of the reconstruction are also obtained with electron spectroscopy.
Misfit between the reconstructed surface layers and the SrTiO$_3$ substrate buried beneath are directly revealed.
Information from the experimental framework is used to design an initial model of the surface reconstructions.
Atomic positions from the initial models are relaxed using density functional theory (DFT).
The relaxed positions are then used to simulate STEM images for direct comparison to experiment.
Analysis of this data set reveals structural and chemical details of the complex (110) surface reconstruction, determined directly as a function of temperature and time.

\begin{figure}
\includegraphics[width=89mm]{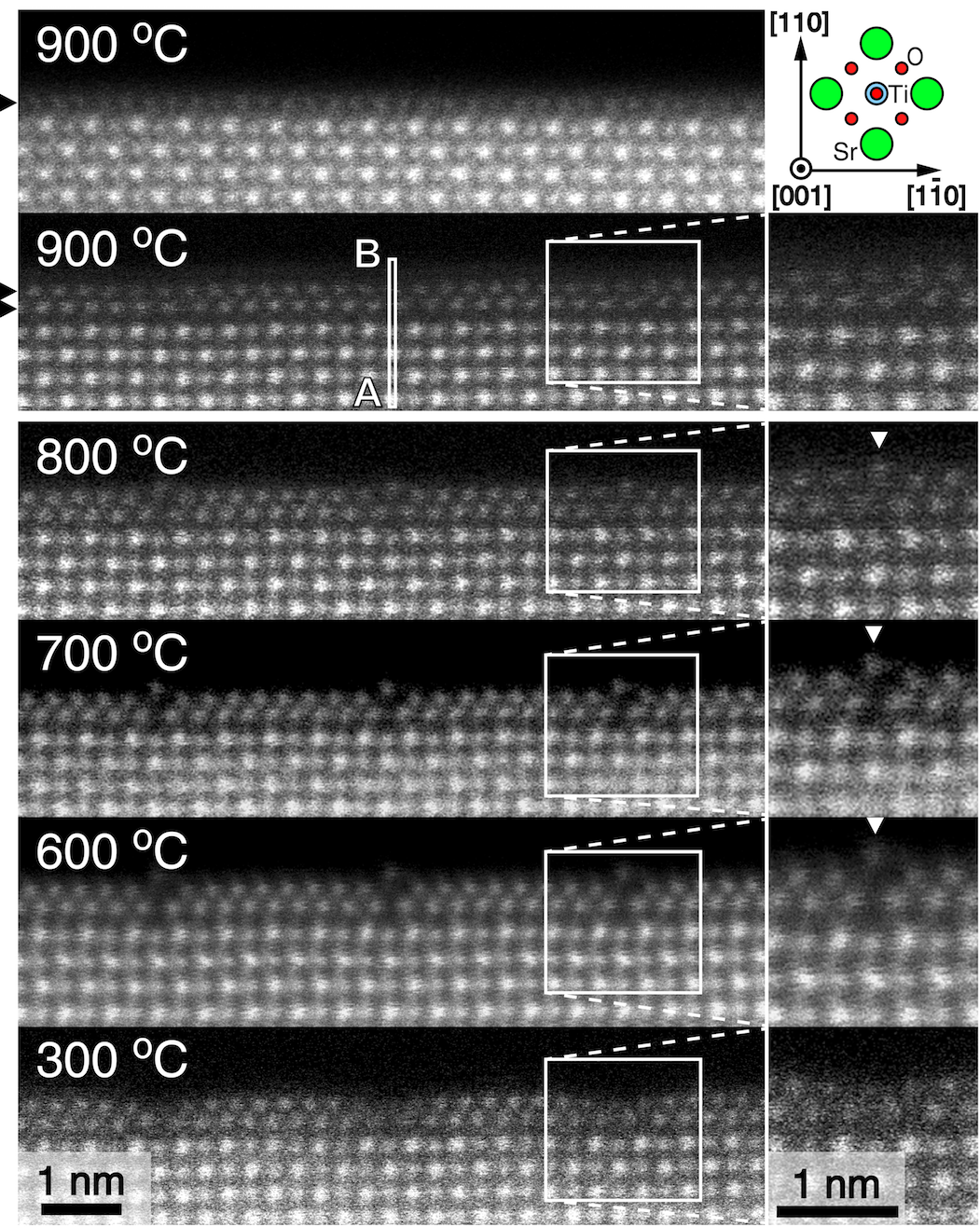}

\caption{ \textbf{left} HAADF-STEM images of reconstructed SrTiO$_3$ (110) surfaces at the indicated temperatures. 
Both single-layer and double layer reconstruction are formed \emph{in-situ} at 900$^{\circ}$C. When cooled to 800$^{\circ}$C, defects appear within the double layer that repeat every 5 Sr-Sr along $\left[1\bar{1}0 \right]$. 
Atoms protrude from the surface, as highlighted by arrows in the corresponding inset \textbf{right}. 
These protruding atoms are lost from the surface at 300$^{\circ}$C.}

\label{fig:temperatureSeries}

\end{figure}

Essential for direct observation of the reconstructed surfaces, thin single crystal specimens are cleanly mounted onto \emph{in-situ} heating devices as detailed in Methods and schematically outlined in Supplementary Figure S1.
Upon heating the bulk SrTiO$_3$ specimen to 900 $^\circ$C in the electron microscope, (110) surfaces are found to reconstruct to either the $4 \times 1$ single layer \cite{LD_Marks_nature_2010} or a double layer reconstruction (Figure \ref{fig:temperatureSeries}). 
We denote the reconstruction by $n \times m$ Wood's notation \cite{Russell_PRB2008}, where $n$ is the repeat unit along $\left[001\right]$, $m$ and is repeat along $\left[1\bar{1}0\right]$.
After prolonged heating, the double layer reconstruction transforms to the $4 \times 1$, see Supplementary Figure S3. 
Furthermore, we also observe that the $4 \times 1$ reconstruction can  transition directly from $1 \times 1$, as shown Supplementary Figure S3.

Recent work by Wang \emph{et al.} has identified the existence of a such a double layer reconstruction in SrTiO$_3$ \cite{Wang2016} in \emph{ex-situ} annealed samples through a combination of STM, DFT, and electron diffraction.
As STM provides real space information about the charge density at the outer layer of the reconstruction, electron diffraction was needed to infer the atom positions, particularly for the second layer.
In contrast, the cross-sectional STEM imaging approach used here provides a route to simultaneously investigate the structure and chemistry of the (110) SrTiO$_3$ surface, and near surface structure, in relation to the bulk.

\begin{figure}
\includegraphics[width=89mm]{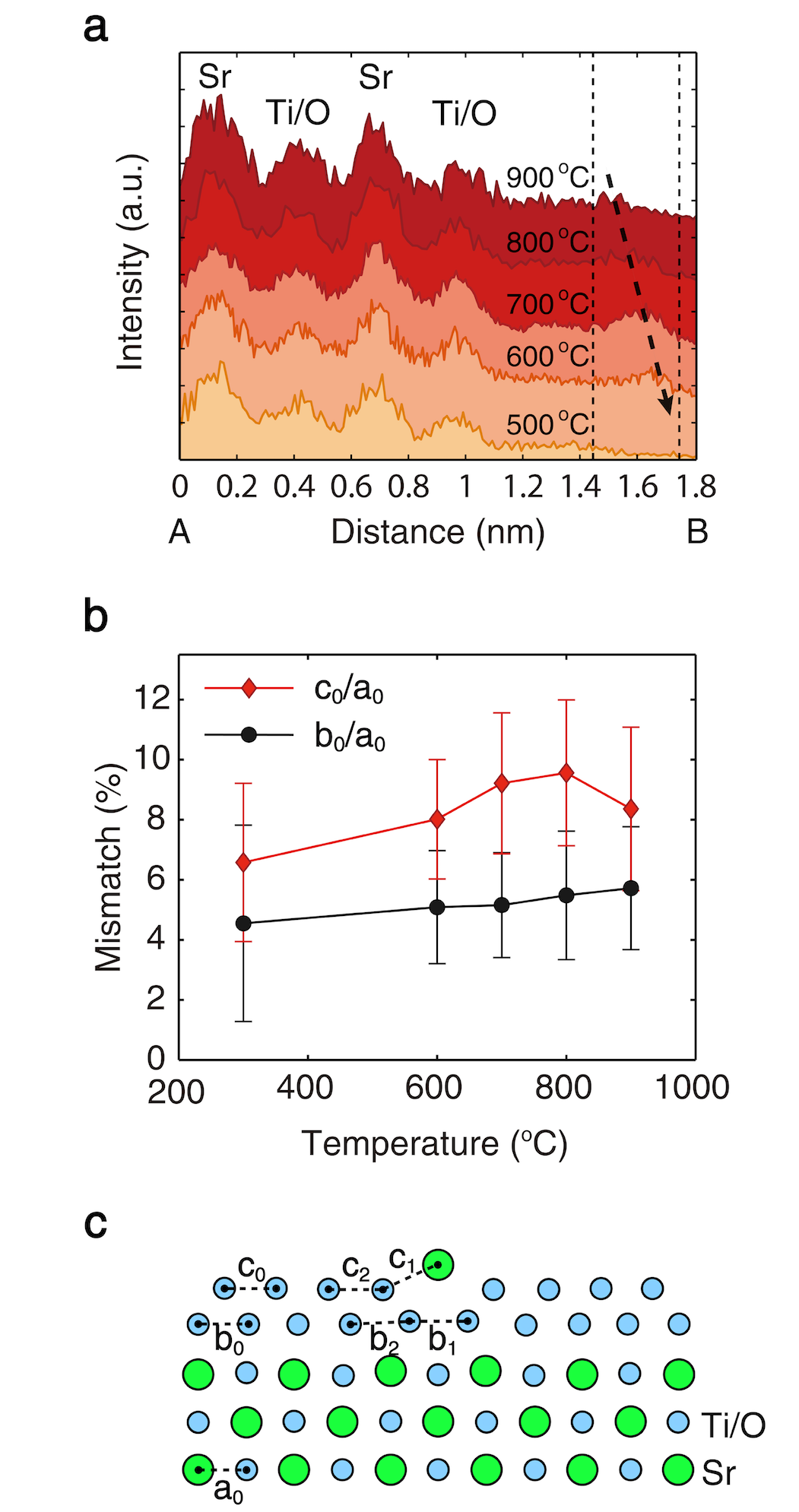}
\caption{
 \textbf{a} Intensity profiles extracted from the  double layer reconstruction at different temperatures along the path A-B defined in Figure 1.
The arrow indicates the protruding atom column.
 \textbf{b} Lattice mismatch between surface layers and the substrate as a function of temperature.
 \textbf{c} The distances used to calculate mismatch:  a$_0$
is the lateral spacing between Sr-Ti/O atom columns in the substrate while b$_x$ and c$_x$ represent the indicated distances in the first and second layer of the reconstructed surface, respectively.}
\label{fig:figure_profile}
\end{figure}

At 900 $^\circ$C, the double layer structure repeats every 4.5-5.5 SrTiO$_3$ unit cells along $\left[1\bar{1}0\right]$ direction, as evidenced by intensity variations of the images shown in Figure \ref{fig:temperatureSeries} and Supplementary Figure S4.
We note that at this temperature, the repeat units of the reconstruction move  dynamically across the surface, as presented in the Supplementary Video 1.
When the sample is cooled to 800 $^\circ$C, the double layer surface reconstruction becomes kinetically stabilized. Furthermore, periodic structures reminiscent of misfit dislocations occur every 5 SrTiO$_3$ unit cells.

To quantify the mismatch between the reconstruction and the substrate, \emph{in-situ} cross-sectional STEM provides direct measurement of lateral distance between atoms of the double layer and substrate.
The relevant distances are shown schematically in (Figure \ref{fig:figure_profile}c).
When the sample is initially cooled from 900$^\circ$C to 800$^\circ$C, the mismatch between the substrate and the outer layer ($(c_0-a_0)/c_0$), increases from 8.4 to 9.6\%. Cooling from 800 $^\circ$C to 300 $^\circ$C, the $c_0$ mismatch with  substrate decreases to 6.6\%. Mismatch for the interfacial layer, $b_0$, is comparatively smaller as it is constrained by the SrTiO$_3$ substrate.

A coincident-site-lattice model suggests that dislocations should form with a repeat distance of about 5 SrTiO$_3$ unit cells at 800$^\circ$C, in excellent agreement with the periodic defects in Figure \ref{fig:temperatureSeries}.
It should be noted that while the repeated structures are not truly dislocation cores as they are only half formed, for convenience, we will refer to them as dislocation nuclei.
Quantification of the spacings near the apparent dislocation nuclei, $b_1$,  finds that they are 10-23\% larger than $b_0$ whereas the neighboring atom column spacing, $b_2$, is about 5-15\% larger than $b_0$.
These measurements indicate significant relaxation surrounding the dislocation nuclei.
The driving force for the temperature dependent structural change is, thus,  due to strain from thermal mismatch.
The repeat structure remains every 5 SrTiO$_3$ unit cells, indicating that it has been ``frozen in'' at low temperatures.

As the double layer reconstruction periodicity stabilizes, we also find additional features of the atomic structure evolution as a function of temperature.
At 800 $^\circ$C atoms begin to protrude from the surface, as marked by arrows in Figure \ref{fig:temperatureSeries} and highlighted in the corresponding inset.
Initially, atoms are displaced by $\sim$90 pm from the topmost surface layer, see Figure \ref{fig:figure_profile}a.
As the sample temperature is reduced to 600$^\circ$C, the protruding atoms move further from the surface to $\sim$140 pm.
At 300$^\circ$C, however, the atoms are readily displaced under the electron probe and can no longer be observed.

\begin{figure}
\includegraphics[width=89mm]{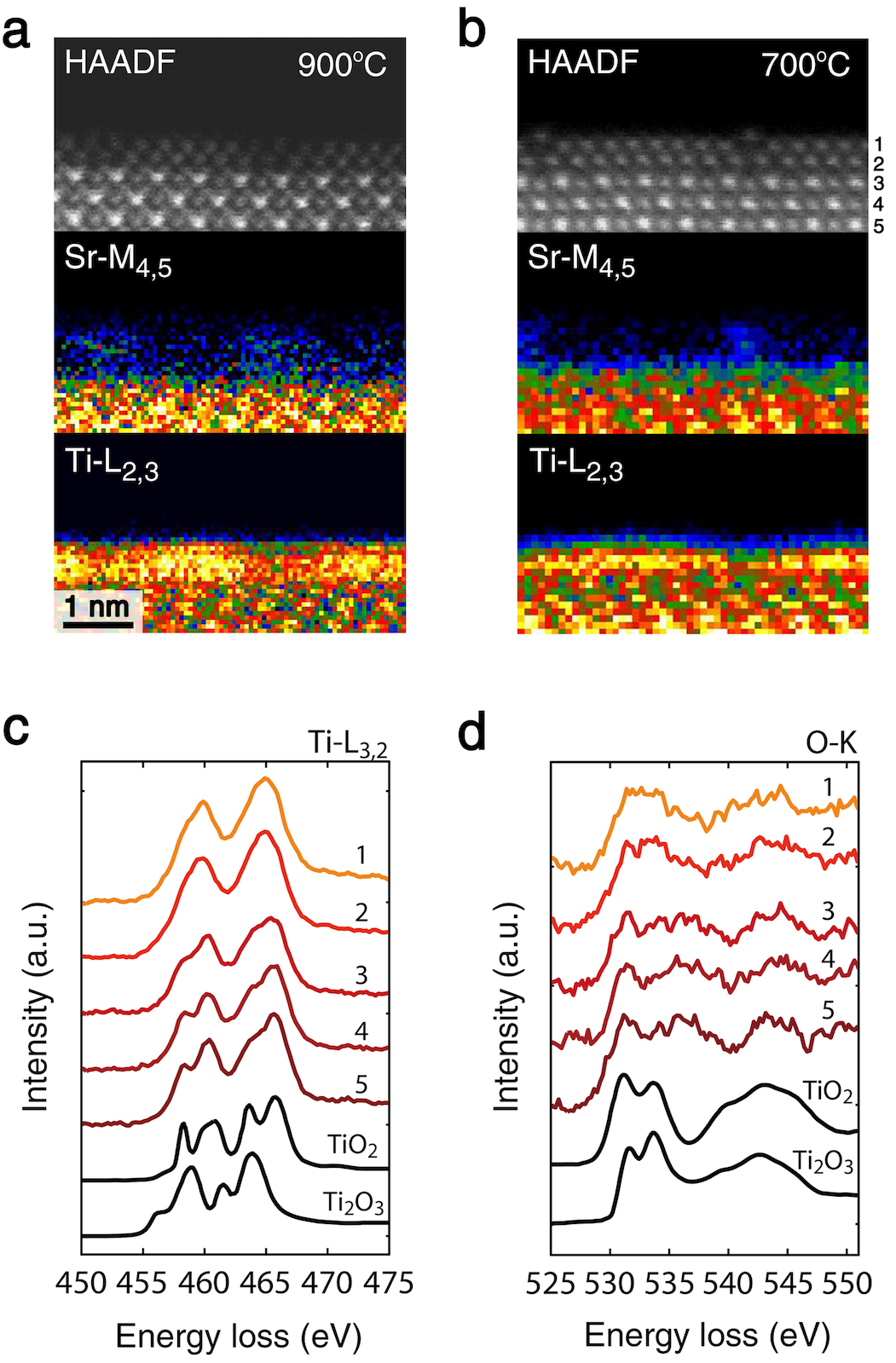}
\caption{EELS elemental mapping of the Sr-L edge and the Ti L$_{3,2}$-edge at  \textbf{a} 900$^{\circ}$C and \textbf{b} 700$^{\circ}$C.
Ti dominates the reconstructed surface, while Sr is localized at the atom  columns protruding from the dislocation nuclei.
HAADF-STEM images of corresponding areas are also shown where the first five surface layers have been labeled.
\textbf{c} Spectra of Ti L$_{3,2}$-edge and \textbf{d} O K-edge at 700$^{\circ}$C for each layer (1-5 in \textbf{b}). Reference spectra of Ti L$_{3,2}$-edge and O K-edge from TiO$_2$ and Ti$_2$O$_3$ are inserted for comparison.\cite{EELS_TixOy}
A change in the fine structure and intensity ratio occurs between layers 1-2 and 3-5, suggesting a mixture of Ti$^{+3}$ and Ti$^{+4}$ in reconstruction .}
\label{fig:EELS}
\end{figure}

A unique aspect of \emph{in-situ} cross-sectional STEM is that it provides a route to directly measure the chemistry of both reconstruction layers at the atomic scale.
Figure \ref{fig:EELS}a shows the distribution of Ti and Sr in the reconstructed double layer at 900$^\circ$C and 700$^\circ$C.
First, the Ti L$_{3,2}$ edge indicates the surface reconstruction is Ti dominated, which is consistent with prior reports \cite{Russell_PRB2008, LD_Marks_nature_2010}.
Second, the Ti distribution is found to be non-uniform in the double layer, varying with the periodic structure seen in HAADF-STEM.
In particular, the Ti appears weakest at the dislocation cores, suggesting Ti deficiency at these locations.
Further, combining the energy loss fine structure of Ti and O (Figure \ref{fig:EELS}c), shows that while Ti is ${+4}$ in the substrate Ti$^{+3}$ is found in the double layer \cite{EELS_TixOy}. 
 Atomic resolution EELS along $\left[1\bar{1}0\right]$ direction also confirms the existence of Ti$^{+3}$ in the double layer (Supplementary Figure S5).
The formation of Ti$^{+3}$ aids in reducing the excess charge of the (110) surface \cite{LD_Marks_nature_2010}.

In contrast to Ti, Sr is detected within the areas of the misfit dislocation cores at 900$^\circ$C. At 800$^\circ$C and below, the signal from Sr becomes localized at the atoms protruding from the top surface.
Because HAADF-STEM is sensitive to the type and number of atoms present in an atom column, the Sr concentration can be estimated from the image intensity.
Using the Sr/Ti intensity ratio from the substrate as a reference, the occupancy of Sr on the top surface is estimated to be only about half of the occupancy of Ti in the neighboring columns.

\begin{figure}
\includegraphics[width=89mm]{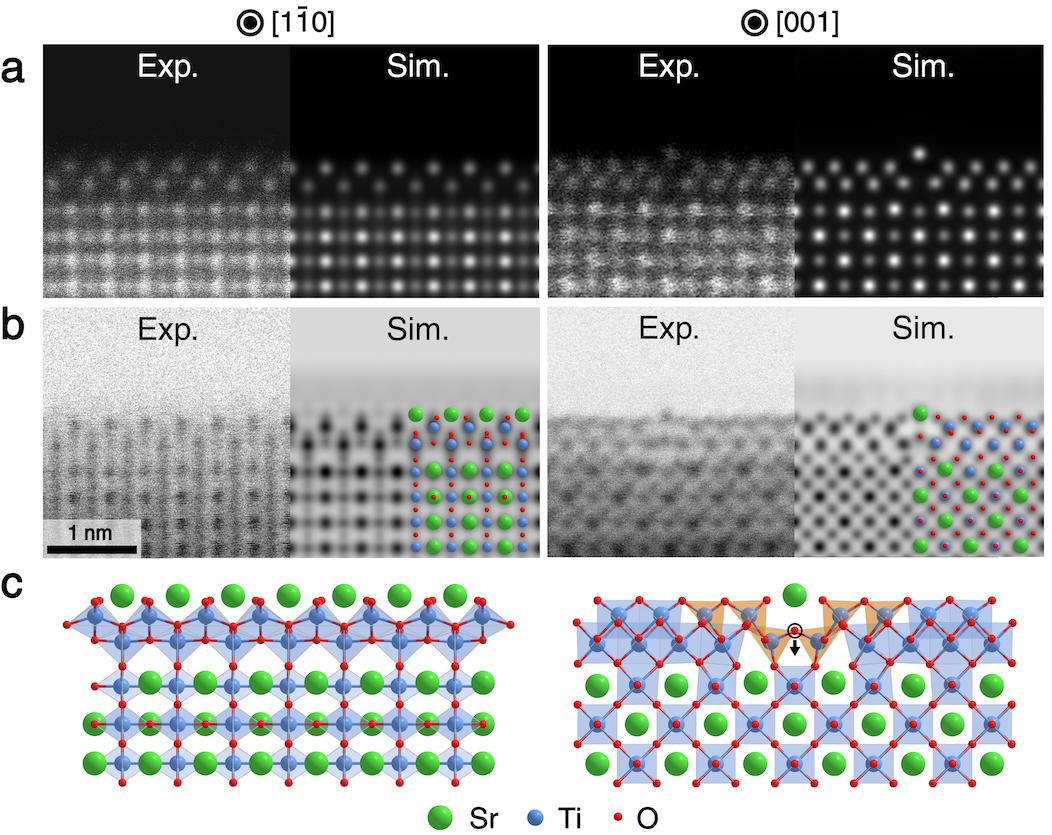}
\caption{Comparison between experiment and image simulations. \textbf{a}, HAADF- and \textbf{b} ABF-STEM images of the reconstructed surface viewed along $\left[1\bar{1}0\right]$ (left) $\left[001\right]$ (right) at 700$^{\circ}$C. Very good agreement is achieved between experimental HAADF, ABF-STEM,  and simulated images.\textbf{c} The corresponding structure model of double layer reconstructed (110) surface. TiO$_5$[] structural units are indicated by orange polyhedra.
}
\label{fig:simulation}
\end{figure}

The accommodation of lattice mismatch in the double layer also influences the oxygen anions.
Using ABF-STEM, the location of oxygen atoms in the double layer structure are also determined at temperatures up to 850$^{\circ}$C (Figure \ref{fig:simulation} and Supplementary Figure S6).
At each temperature, we find a reconstructed surface terminated by oxygen.
Based on these images, plane of oxygen at the surface is closer to Ti than those in the first layer of the reconstruction.
Furthermore, this indicates that Ti-O octahedral bonding is maintained in double layer.
Such octahedral bonding is widely seen in Ti oxide polymorphs, and is in agreement with a recent model of the $2 \times 5$a reconstruction \cite{Wang2016}.
As the surface is terminated by negatively charged oxygen, Sr$^{2+}$ within the free volume of the dislocation nuclei aids in accommodating excess charge.

Based on imaging and spectroscopy along $\left[001\right]$ and $\left[1\bar{1}0\right]$ (Figure \ref{fig:simulation}a and \ref{fig:simulation}b), a three dimensional model of the double layer structure is directly constructed, and then structurally refined via DFT calculations (Figure \ref{fig:simulation}c).
From the model, important structural characteristics are revealed. 
First, at the dislocation nuclei, Ti atoms are surrounded by only five nearest neighbor oxygen.
As a result, TiO$_5$[] units are formed at the dislocation nuclei, as indicated by the orange polyhedra. 
Second, oxygen in the middle of dislocation nuclei also shift downward (arrow in Figure \ref{fig:simulation}c) as they try to maintain the Ti-O bond length.
Third, the refined structure shows some similarities to the ($2\times5$)a SrTiO$_3$ (110) surface reconstruction proposed by Wang et al.\cite{Wang2016}, though the column density of protruding Sr and interfacial layer Ti are different.
Finally, using the DFT refined structure, HAADF and ABF-STEM images have been simulated.  
Both are in excellent agreement with the experimental results ( Figures \ref{fig:simulation}a,b).
Simulated HAADF and ABF-STEM images determined using the $2\times5$a reconstruction geometries published by Wang et al.\cite{Wang2016}, however, exhibit discrepancies with our STEM observations as highlighted in Supplementary Figure S7.
Overall, these results point to the power of \emph{in-situ} cross-sectional STEM measurements to provide new, complementary surface reconstruction details.

Beyond surface reconstructions, the approach applied here provides insight into the growth of thin films. 
Specifically, the reconstructed double layer naturally serves as a template for growth of additional Ti oxide layers.
For example, Supplementary Figure S8 shows that multiple layers of titanium oxide can grow epitaxially on the reconstructed double layer surface.
As growth proceeds, the dislocations become fully formed, where the overgrown thin film is coherent with double layer.
Without the cross-sectional geometry, the complexities of this structure would otherwise be lost, which would occur for conventional surface-only characterization techniques.

Here we have demonstrated that STEM is capable of measuring both the reconstructed surface and \emph{near-surface} structure of large scale single crystals directly in real-space for the first time.
Applied to (110) SrTiO$_3$, a variety of environmental conditions can be employed to observe the time evolution of the structure.
Furthermore, coupling the approach to EELS provides a unique ability to probe the evolution of the atomic structure while at the same time assessing changes in chemistry and surface oxidation states.
For the first time, \emph{in-situ} atomic scale imaging of single crystals can provide a route to directly explore reconstructed surfaces, serving as a direct input to first principles calculations to determine the origins of stable surface reconstructions.

\section{Methods}
\subsection{Sample preparation.} 
\label{sub:samplePrep}


 Single crystals of SrTiO$_3$ (MTI Corporation) with surface normals of [100] and [110] were first wedge-polished in cross-section with an Allied MulitPrep\texttrademark\ system.
 Subsequently, the wedge shaped samples were affixed to a diamond TEM grid (SPI Supplies).
 The samples were then thinned to electron transparency with a Fischione 1050 argon ion mill, ending with a final cleaning step at 0.5 keV.
 The as-milled samples were then fractured into pieces with sizes ranging from 20 to 100 $\mu$m.
 A glass needle was then used to select and lift-out thin regions from amongst the fractured debris, and transferred onto a Protochips heating membrane.
 The success of transfer arises from the static and van der Waals forces between the small single crystal pieces and the glass needle, which is commonly employed in the focused ion beam ex-situ lift-out method \cite{Liftout_2015}.
 A schematic overview of the process is shown in Supplementary Figure S1.

\subsection{In-situ heating and characterization.} 
\label{sub:insituDetails}


 \emph{In-situ} observations were performed using a probe-corrected FEI Titan G2 microscope operated at 200 kV and equipped with a Gatan Enfinium spectrometer.
 The electron probe current was measured with the screen to be approximately 40 pA with a probe convergence semi-angle of 19.6 mrad.
 The inner detector collection semi-angle for HAADF-STEM imaging was approximately 77 mrad while for ABF imaging the inner and outer detector collection semi-angles were approximately 7 and 20 mrad, respectively.
 A Protochips Aduro double tilt holder was used to heat the membrane to 900 $^\circ$C at ad rate of 1 $^\circ$C/s.
 The specimen was then equilibrated at 900 $^\circ$C for about 90 minutes before cooling to the following temperatures: 800 $^\circ$C, 700 $^\circ$C, 600 $^\circ$C, and finally 300 $^\circ$C.
 At each step, the cooling rate was 1$^\circ$C/s.
 Electron energy loss spectra were acquired with an energy dispersion of 0.25 eV, a collection semi-angle of 39 mrad, and an energy resolution of 1 eV.

\subsection{Structure quantification and image simulation.} 
\label{sub:structureQuant}


 Image analysis was performed using custom MATLAB scripts with the Image Processing Toolbox.
 The atom column positions in the HAADF-STEM images were measured by finding the normalized cross-correlation with a 2D Gaussian template \cite{Sang2014a}.
 Intensity line profiles were extracted from the unprocessed, original image with an integration width of about 0.1 nm.
 Simulations for the HAADF and ABF-STEM images were conducted using the multislice algorithm. 
Crystal and probe wavefuntions were sampled at a minimum of 2.3 pm/pix, and probe positions were chosen to satisfy the Nyquist sampling theorem. 
Simulated images were resampled with Fourier interpolation and then blurred with 0.1 nm full-width at half-maximum Gaussian to account for the effects of a finite effective size of the electron source \cite{LeBeau2008}.
Where indicated, the simulated model was either derived directly from the experimental observations or from Ref.~\citenum{Wang2016}. 

\subsection{First principles calculations} 
\label{sub:DFT}

Relaxation of predicted surface structures were performed in VASP using the projector-augmented-wave (PAW) method \cite{VASP_1} with a plane-wave cutoff of 520 eV in the PBE-GGA scheme \cite{VASP_GGA}.
The PAW potentials contained 10, 12, and 6 valence electrons for Sr (4s$^2$4p$^6$5s$^2$), Ti (3s$^2$3p$^6$4s$^2$3d$^2$), and O (2s$^2$2p$^4$) respectively.
Spin polarization was accounted for in all calculations.
Ions not fixed in the initial structures detailed below were allowed to relax until all forces were less than 0.01 eV/$\AA$.
Under this set of parameters, the lattice parameter of bulk SrTiO$_3$ was calculated as 3.942 $\AA$ which is in agreement with the experimental value of 3.905 $\AA$ at room temperature. A $5 \times 5 \times 5$ and $5 \times 1 \times 1$  k-point mesh was used for the bulk primitive cell and $2 \times 5$ cell, respectively.

A slab model for the double layer reconstruction was derived from \emph{in-situ} images of the surface profiles in Fig. \ref{fig:temperatureSeries}.
In addition to the reconstructed surface, the model included 3 SrTiO$^{4+}$ and 4 O$_2^{4-}$ layers of which 1 and 2 layers respectively were fixed at the geometry calculated for the bulk.
10 $\AA$ of vacuum separated opposite sides of the slab model.
Upon reaching the specified convergence criteria, the relaxed structures were used to generate simulated HAADF and ABF-STEM images to evaluate agreement with the \emph{in-situ} images.


\bibliographystyle{naturemag}
\bibliography{refs.bib}

\section{Acknowledgements}


Authors gratefully acknowledge support from the National Science Foundation (NSF) (Award No. DMR-1350273).
E.D.G.~acknowledges support for this work through a National Science Foundation Graduate Research Fellowship (Grant DGE-1252376).
D.L.I.~and P.C.B.~gratefully acknowledge support for this work from the NSF under DMR-1151568 and the Air Force Office of Scientific Research under contract FA9550-14-1-0264.
Authors acknowledge the use of the Analytical Instrumentation Facility (AIF) at North Carolina State University, which is supported by the State of North Carolina and the National Science Foundation.

\section{Author Contributions}

W.X.~performed the in-situ sample preparation, experiment and data analysis. P.C.B.~and D.L.I.~performed the first principle calculations. E.D.G.~performed the STEM multislice simulations. J.M.L.~supervised the work. J.M.L.~and W.X.~co-wrote the initial draft.   All authors contributed to discussing, revising, and editing the manuscript.

\section{Correspondence}

Correspondence and requests for materials
should be addressed to J.M.L.~(email: jmlebeau@ncsu.edu).

\section{Competing Financial Interests}

The authors declare no competing financial interests.

\end{document}